\documentclass[%
 reprint,
 amsmath,amssymb,aps,
]{revtex4-2}

\usepackage{graphicx}
\usepackage{dcolumn}%
\usepackage{bm}
\usepackage{hyperref}

\begin{document}
\title{Gravity Induced CP Violation}
\author{J-M Rax}
\email{jean-marcel.rax@universite-paris-saclay.fr}
\affiliation{IJCLab-UMR9012-IN2P3 \\Universit\'{e} de Paris-Saclay\\  
 Facult\'{e} des Sciences d'Orsay\\
91405 Orsay France}
\date{\today}

\begin{abstract}
The impact of earth's gravity on neutral kaons oscillations is analyzed. The
main effect of a Newtonian potential is to couple the strangeness
oscillation and the strange quarks zitterbewegung. This
coupling is responsible for the observed $CP$\ violations. Gravity induced $
CP$ violation is in fact a $CPT$\ violation with $T$\ conservation rather
than a $T$\ violation with $CPT$\ conservation, but the finite lifetime of
the short-lived kaons induces a rotation of the imaginary $CPT$ violation
parameter such that it becomes real and the effect is observed as a $CP$ and
a $T$ $\ $violation. Both indirect and direct violation parameters 
are predicted in agreement with the experimental data.
\end{abstract}

\maketitle

\section{Introduction}

The standard model (SM) is an experimentally well-tested theory describing
elementary particles dynamics. The small $CP$ violation (CPV) measured in
neutral kaons experiments is satisfactorily described within the SM
framework by the Cabibbo-Kobayashi-Maskawa (CKM) matrix \cite{1,2}.
However the ultimate origin of CPV remains an open issue.

In this study we consider the kaon-antikaon oscillations and we demonstrate
that the free parameter involved in the construction of the CKM matrix to
describe CPV is ultimately due to a small coupling, induced by earth's
gravity, between the strange quarks zitterbewegung oscillations and the
kaons mixing oscillations $K^{0}\rightleftharpoons \overline{K}^{0}$.

The fact that gravity is the ultimate source of CPV opens very interesting
perspectives to set up cosmological models with asymmetric baryogenesis
compatible with the present state of our universe.

As $K^{0}/\overline{K}^{0}$ oscillations is a very low energy process ($
10^{-6}$ eV) there are no needs to rely on quantum field theory (QFT) and we
can use the well established Lee, Oehme and Yang (LOY) effective Hamiltonian 
model \cite{3}. This effective Hamiltonian offers the appropriate framework
to discuss discrete symmetries conservation or violation \cite{4}.

The LOY model is completed here with the usual Newtonian gravitational term
describing the influence of earth's gravity on kaons dynamics. The final
results obtained after a careful analysis of this first principles effective
Hamiltonian model (LOY + Newtonian gravity) agree with the experimental
results: both CPV observable parameters $
\mathop{\rm Re}
\varepsilon $ and $
\mathop{\rm Re}
\varepsilon ^{\prime }/\varepsilon $ are predicted in agreement with the
experimental data. Moreover, with gravity induced CPV there is no $T$
violation at the microscopic level. The observed $T$ violation stems from
the irreversible decay of the short lived kaons $K_{S}$ continuously
regenerated from the long-lived one $K_{L}$ by the (gravity induced) coupling between 
$\Delta S=2$ oscillations and quarks zitterbewegung.

The study of discrete symmetries breaking by the fundamental interactions
was initiated by the discovery of $P$ violation by C. S. Wu et al. in 1957 
\cite{5} following the Lee and Yang suggestion of $P$ non conservation in
weak interactions \cite{6}. After their study of the $P$ symmetry, Lee, Yang
and Oehme set up the canonical LOY model \cite{3}, completed later by T. T.
Wu and Yang \cite{7}, to understand the interplay between $C$, $P$ and $T$
symmetries and describe the dynamics of neutral kaons. Sixty years ago,
Christenson, Cronin, Fitch and Turlay reported the first measurement of CPV
through the observation of long-lived neutral kaons $K_{L}$ decays into two
charged pions, a decay forbidden by $CP$ conservation \cite{8}. 

Understanding the ultimate origin of this observed CPV is of prime
importance because, as suggested by Sakharov \cite{9}, it might explain how
our matter-dominated universe emerged during its early evolution. 

The observed present level of CPV is far too small to build cosmological
evolution models compatible with the observed dominance of baryons over
antibaryons in our universe. The gravity induced CPV identified and analyzed
here renews this issue in depth.

Following Good's early analysis \cite{10}, the impact of an external or a
gravitational field on kaons oscillations has been considered in several
studies \cite{11,12,13,14,15,16}. In
particular, in Ref. \cite{17}, E. Fishbach considered the coupling parameter
between earth's gravity and $K^{0}/\overline{K}^{0}$ oscillations: $g\hbar
/2\delta mc^{3}$ where $g$ is the acceleration of gravity and $2\delta m$
the mass difference between $K_{L}$ and $K_{S}$. This coupling parameter is
far too small to explain the observed CPV in $K^{0}/\overline{K}^{0}$
experiments on earth, but Fishbach pointed out the fact that, if this small
parameter is multiplied by the large factor $m/2\delta m$ ($m$ is the
neutral kaon mass), we obtain half the value of $
\mathop{\rm Re}
\varepsilon $ the indirect CPV parameter. We demonstrate here that this was
more than a simple numerical coincidence.

This paper is organized as follows, in the next section we briefly review
the LOY model, an in depth presentation of this canonical model can be found
in \cite{4}. 

The inclusion of the usual Newtonian gravity is performed in
section III. We analyze the impact of this additional term and discover that
it requires the evaluation of the velocity matrix element of the internal
strange quark motion inside the kaons. This evaluation is achieved in
section IV where we consider the zitterbewegung fast motion of a 
fermion-antifermion pair described by Dirac's equation \cite{18}. Section V
is devoted to the solution of the LOY model including Newtonian gravity. The
two CPV parameters $\varepsilon $ and $\varepsilon ^{\prime }$ are predicted
in this section on the basis of this new model and in full agreement with the
experimental data. The last section gives our conclusions.

The study presented below complements a previous study \cite{19} based on two coupled
Klein-Gordon equations on a Schwarzschild metric rather than a
Newtonian framework with  two coupled Schr\"{o}dinger equations used here.
The convergence of the results of this new Newtonian model with the previous
general relativity model is remarkable and gives confidence in the relevance
of the results which are thus model independent.

\section{Neutral kaons oscillations and decays without gravity}

After its production, a generic neutral kaons state $\left| \Psi
\right\rangle $ behaves as a linear superposition of the $\left|
K^{0}\right\rangle $ particle state and\ the $\left| \overline{K}
^{0}\right\rangle $ antiparticle state. In the kaon rest frame, the
amplitudes $a$ and $b$ of this superposition are simply functions of the
kaon proper time $\tau $
\begin{equation}
\left| \Psi \right\rangle =a\left( \tau \right) \left| K^{0}\right\rangle
+b\left( \tau \right) \left| \overline{K}^{0}\right\rangle \text{,}
\label{ampK}
\end{equation}
where $\left\langle K^{0}\right. \left| K^{0}\right\rangle $ $=$ $%
\left\langle \overline{K}^{0}\right. \left| \overline{K}^{0}\right\rangle $ $%
=$ $1$ and $\left\langle \overline{K}^{0}\right. \left| K^{0}\right\rangle
=0 $. 

The particle and antiparticle states, $\left| K^{0}\right\rangle $ and 
$\left| \overline{K}^{0}\right\rangle $, are eigenstates of the strangeness
operator $S$ with eigenvalues $S=\pm 1$. The relative phase between these
two states is not observable and is fixed by convention. In this study we
take the convention $CP\left| K^{0}\right\rangle =\left| \overline{K}%
^{0}\right\rangle $.\ It is to be noted that if we consider the ratio of
amplitudes associated with two decay channels, the phase convention must
be eliminated through an appropriate rephasing factor to obtain meaningful
physical quantities independent of the phase convention. This point is
particularly important to interpret direct CPV experiments. The two
normalized $CP$ eigenstates, with eigenvalues $\mp 1$, are
\begin{equation}
\left| K_{2/1}\right\rangle =\left( \left| K^{0}\right\rangle \mp \left| 
\overline{K}^{0}\right\rangle \right) /\sqrt{2} \text{.} \label{eigen}
\end{equation}
These are also energy eigenstates with masses $ m\pm \delta
m $. 

Without CPV, the long-lived and short-lived kaons $\left|
K_{L}\right\rangle =\left| K_{2}\right\rangle $, $\left| K_{S}\right\rangle
=\left| K_{1}\right\rangle $ and $\left\langle K_{L}\right. \left|
K_{S}\right\rangle $ $=$ $0$. 

Kaons are unstable and $d\left\langle \Psi
\right. \left| \Psi \right\rangle /d\tau <0$. To ensure a non unitary time
evolution, the rest frame LOY effective Hamiltonian operator restricted to
the $\left[ \left| K^{0}\right\rangle ,\left| \overline{K}^{0}\right\rangle
\right] $ Hilbert space is non Hermitian. The Weisskopf-Wigner approximation 
\cite{20} provides a satisfactory description of the decays and is adapted
to construct the effective Hamiltonian ensuring the proper time evolution
and decay of the $\left| \Psi \right\rangle $ state
\begin{equation}
j\hbar \frac{d\left| \Psi \right\rangle }{d\tau }=\left[ mc^{2}\widehat{I}-
\widehat{\delta m}c^{2}-j\hbar \widehat{\gamma }\right] \cdot \left| \Psi
\right\rangle \text{,}  \label{km2}
\end{equation}
where $\widehat{I}$ is the $2\times 2$ identity matrix. The coupling and decay
of the amplitudes $a$ and $b$ are described respectively by the operators $%
\widehat{\delta m}$ and $\widehat{\gamma }$ whose matrix representations on
the $\left[ \left| K^{0}\right\rangle ,\left| \overline{K}^{0}\right\rangle
\right] $ basis are

\begin{equation}
\widehat{\delta m}=\left[ 
\begin{array}{cc}
0 & \delta m \\ 
\delta m & 0
\end{array}
\right] \text{, }\widehat{\gamma }=\frac{1}{2}\left[ 
\begin{array}{cc}
\Gamma & -\delta \Gamma \\ 
-\delta \Gamma & \Gamma
\end{array}
\right] \text{.}  \label{m123}
\end{equation}

The mass and the decay operators matrix elements of this effective LOY
Hamiltonian are related to first and second order weak interaction couplings
with a set of virtual (mass) or final (decay) states $f$ $\ $with energy $
E_{f}$: 
\begin{eqnarray}
m\delta _{i}^{j}+\left\langle i\right| \widehat{\delta m}\left|
j\right\rangle  &=&m_{0}\delta _{i}^{j}+\left\langle i\right| \widehat{H}%
_{wk}\left| j\right\rangle /c^{2}  \nonumber \\
&&+\Sigma _{f}VP\frac{\left\langle i\right| \widehat{H}_{wk}\left|
f\right\rangle \left\langle f\right| \widehat{H}_{wk}\left| j\right\rangle
/c^{2}}{m_{0}c^{2}-E_{f}}\text{,}  \nonumber \\
\hbar \left\langle i\right| \widehat{\gamma }\left| j\right\rangle  &=&\pi
\Sigma _{f}\delta \left( m_{0}c^{2}-E_{f}\right)   \nonumber \\
&&\times \left\langle i\right| \widehat{H}_{wk}\left| f\right\rangle
\left\langle f\right| \widehat{H}_{wk}\left| j\right\rangle \text{,}
\label{mass}
\end{eqnarray}
where $\widehat{H}_{wk}$ is the weak interaction Hamiltonian, $VP$ is the
Cauchy principal value and $O\left[ H_{wk}^{3}\right] $ and higher orders
terms are neglected \cite{4}.

The ordering of (\ref{km2}) and (\ref{m123}) is: ({\it i}) $\delta m/m\sim
3.5\times 10^{-15}$, ({\it ii}) $\Gamma -\delta \Gamma /\Gamma +\delta
\Gamma \sim 600$ and ({\it iii}) $2\delta mc^{2}/\hbar \left| \delta \Gamma
\right| \sim 2\delta mc^{2}/\hbar \Gamma \sim 0.94$ \cite{21}.

The analysis of the impact of earth's gravity on neutral kaons experiments
will be carried out in two steps. First, we restrict the dynamics to stable
particles ($\widehat{\gamma }=\widehat{0}$) interacting with gravity: to do
so we set $\Gamma =\delta \Gamma =0$ in (\ref{km2}, \ref{m123})
and complete the model with the small perturbation associated with Newtonian
gravity. Then, we set up a careful analysis of the impact of the finite
lifetime of the $K_{S}$ ($\Gamma \neq 0$, $\delta \Gamma \neq 0$) on the
predicted gravity induced anomalous $K_{S}$ regeneration. 

The result of this
two steps analysis is that : ({\it i}) at the theoretical level gravity
induces a small $CPT$ violation on stable particles dynamics and $T$ is 
preserved, and ({\it ii}) at the experimental
level, because of the finite $K_{S}$ lifetime, this violation is observed
and measured as an apparent $CP$ violation and $T$ violation. This very
critical point led in the past to assume $CPT$ conservation. Among the
consequences of this assumption, the repahasing factor involved in direct
CPV analysis was taken equal to one. We demonstrate here that this rephasing
factor is different from one and that it provides a new interpretation of
the observed direct CPV.

The $C$, $P$, $T$ symmetries, conservation or violation, are  usually
identified as follows in a perturbed LOY model \cite{4}: ({\it i}) if in (\ref{km2}) 
and (\ref{m123}) gravity induces a perturbation such that $
m\rightarrow m\pm j\rho $, where $\rho $ has a small real part, then $T$ is
conserved, and ({\it ii}) if in (\ref{km2}) and (\ref{m123}) gravity induces
a perturbation such that $\delta m\rightarrow \delta m\pm \rho $, where $
\rho $ has a small real part, then $CPT$ is conserved. In both cases $CP$
is not conserved. Gravity induces a type ({\it i}) perturbation, but the $
K_{S}$ finite lifetime make it observed as a type ({\it ii}) perturbation.

The $CPT$ theorem is usually demonstrated on the basis of three main
assumptions: ({\it i}) Lorentz group invariance, ({\it ii}) spin-statistics
relations and ({\it iii}) local field theory. In the rest frame of a meson
interacting with a massive spherical object like earth the assumption ({\it i%
}) is not satisfied so we must not be surprized that the $CPT$ theorem no
longer holds.

\section{Impact of Newtonian gravity on neutral kaons oscillations}

The rest frame Schr\"{o}dinger equation (\ref{km2}) 
and (\ref{m123}) for stable kaons ($\widehat{\gamma }%
=\widehat{0}$) is based on the assumption that the rest frame energy of 
the uncoupled particles with mass $m$ is $E=mc^{2}$. On the surface of a
spherical massive object like earth (mass $M_{\oplus }$ $=$ $5.9724\times
10^{24}$ kg, radius $R_{\oplus }$ $=$ $6.3781\times 10^{6}$ m), the rest
frame energy $E$ of a particle with mass $m$ and radial position $r$ must be
completed with the Newtonian potential energy
\begin{equation}
E=mc^{2}-G_{N}M_{\oplus }m/r\text{,} \label{energy}
\end{equation}
where $G_{N}$ is the constant of gravitation and $G_{N}M_{\oplus
}m/R_{\oplus }$ $\sim 10^{-9}mc^{2}$.
In a CPV kaon experiment $r$ is the instantaneous barycentric position of
the mass/energy associated with the $d/\overline{d}$ and $s/\overline{s}$
quarks dynamics inside the $K_{S}$ and $K_{L}$ as $\left| K^{0}\right\rangle
=$ $\left| d\overline{s}\right\rangle $ and $\left| \overline{K}%
^{0}\right\rangle =\left| \overline{d}s\right\rangle $. 
This instantaneous position $r$ is the sum of an average position plus a fast oscillation of
the $d/\overline{d}$ and $s/\overline{s}$ quarks inside the kaons.

In the
earth laboratory frame the average center of mass/energy follows a very slow
(undetectable) vertical free fall combined with a fast horizontal inertial
motion almost parallel to the surface of earth. The rest frame of the kaons
follows this average motion.

In addition to this average motion, inside the kaons, quarks perform an
unknown, small scale, high frequency, fast motion with respect to the
average center of mass/energy. The internal kinetic and potential (strong
interaction) energies associated with this internal dynamics of the bound
quarks ($K^{0}\sim d-\overline{s}$, $\overline{K}^{0}\sim \overline{d}-s$)
is part of the mass term $m_{0}$ in (\ref{mass}), but both ({\it i}) the
average and ({\it ii}) the small fluctuations (due to the quarks motions) of the Newtonian potential are not taken
into account in the rest frame LOY Hamiltonian (\ref{km2}) and (\ref{m123}).

We consider this small additional Newtonian contribution to the kaon
dynamics as follows. The radial position $r$ involved in the rest frame
energy (\ref{energy}) is decomposed as $r=R_{\oplus }+X+x$ the sum of: ({\it %
i}) the earth radius $R_{\oplus }$, plus ({\it ii}) the average slow
vertical kaon displacement $X\left( \tau \right) $ with respect to $%
R_{\oplus }$, plus ({\it iii}) the unknown, high frequency, fast internal
vertical motion $x$, around the average vertical position $R_{\oplus }+X$,
associated with the $d/\overline{d}$ and $s/\overline{s}$ quarks dynamics.

The ordering $X+x\ll R_{\oplus }$ allows to expand the kaon potential energy
(\ref{energy}) as 
\begin{equation}
E=mc^{2}-G_{N}M_{\oplus }m/R_{\oplus }+mgX+mgx\text{,}  \label{enerfgy2}
\end{equation}
where $g=G_{N}M_{\oplus }/R_{\oplus }^{2}=9.806$ m/s$^{2}$ is the
acceleration of gravity on earth. The proper time $\tau $ is the kaon rest
frame time. This rest frame follows the average vertical motion $X\left(
\tau \right) $. Thus $X$ does not operate kinematically on the kaon state: $%
mgX$ is not observable as a kaon dynamical variable in the kaon rest frame,
it is just an additional time dependant (negligible) energy.

The case is different for the quarks internal vertical motions $x$, around $%
R_{\oplus }+X$, because $\tau $ is not the proper time of the quarks. The
internal position operator $\widehat{x}$ operates in the tensorial product
space of the $d/\overline{d}$ and $s/\overline{s}$ Dirac's spinors states.
To complete the LOY model with a small gravitational potential energy, we
introduce the representation of the vertical position operator $\widehat{x}$
in the kaons Hilbert space $\left[ \left| K^{0}\right\rangle ,\left| 
\overline{K}^{0}\right\rangle \right] $ as 
\begin{eqnarray}
    \widehat{x} &=&\left\langle d\overline{s}\right| \widehat{x}\left| d%
    \overline{s}\right\rangle \left| K^{0}\right\rangle \left\langle
    K^{0}\right| +\left\langle \overline{d}s\right| \widehat{x}\left| \overline{d%
    }s\right\rangle \left| \overline{K}^{0}\right\rangle \left\langle \overline{K%
    }^{0}\right|   \label{xx} \\
    &&+\left\langle \overline{d}s\right| \widehat{x}\left| d\overline{s}%
    \right\rangle \left| \overline{K}^{0}\right\rangle \left\langle K^{0}\right|
    +\left\langle d\overline{s}\right| \widehat{x}\left| \overline{d}%
    s\right\rangle \left| K^{0}\right\rangle \left\langle \overline{K}%
    ^{0}\right| \text{.}  \nonumber
    \end{eqnarray}

The operator $\widehat{x}$ in (\ref{xx}) is the time dependant (Heisenberg
representation) position operator and the diquark states $\left| q\overline{%
q^{\prime }}\right\rangle $ are $q$ and $\overline{q^{\prime }}$ quarks  
states in the effective confining potential of a kaon. In Ref. \cite{19}, 
within the framework of general relativity, we have used
the Schr\"{o}dinger representation.

We neglect the dynamics of the $d/\overline{d}$ quark component as the
ratio of the mass of the $d$ quark over the mass of the $s$ quark is of the
order of $0.04$. The results will therefore have a good relative accuracy of 4\%.
The instantaneous motion of the mass/energy-barycenter
inside a kaon is dominated by the $s/\overline{s}$ dynamics.
The lighter quarks $d$ and $\overline{d}$ are just followers of the kaons average motion
and passive witness of the $s/\overline{s}$ (zitterbewegung) fast motion.

The position operator to be used in the kaon rest frame Hilbert space $
\left[ \left| K^{0}\right\rangle ,\left| \overline{K}^{0}\right\rangle
\right] $ is thus reduced to 
\begin{eqnarray}
\left. \widehat{x}\right| _{LOY} &\equiv &\left\langle \overline{s}\right| 
\widehat{x}\left| \overline{s}\right\rangle \left| K^{0}\right\rangle
\left\langle K^{0}\right| +\left\langle s\right| \widehat{x}\left|
s\right\rangle \left| \overline{K}^{0}\right\rangle \left\langle \overline{K}%
^{0}\right|   \nonumber \\
&&+\left\langle s\right| \widehat{x}\left| \overline{s}\right\rangle \left| 
\overline{K}^{0}\right\rangle \left\langle K^{0}\right| +\left\langle 
\overline{s}\right| \widehat{x}\left| s\right\rangle \left|
K^{0}\right\rangle \left\langle \overline{K}^{0}\right| \text{.}  \nonumber
\\
&&  \label{xs}
\end{eqnarray}
In the following calculations we will use also the velocity operator 
$d\widehat{x}/d\tau $, given by Heisenberg's equation
 \begin{equation}
    j\hbar d\widehat{x}/d\tau =\widehat{x}\cdot \widehat{H}_{sK}-\widehat{H}%
    _{sK}\cdot \widehat{x}\text{.}
    \end{equation}
 where $\widehat{H}_{sK}$ is the effective Hamiltonian
describing the strange quarks confinement inside the kaon.

We will demonstrate that the velocity operator $d\widehat{x}/d\tau $
fulfilling this relation is in fact independent of the mass, charge and
potential involved in $\widehat{H}_{sK}$. 

The generic name {\it %
zitterbewegung} was given to this nonintuitive result demonstrated in
section IV \cite{18}. 

Within the framework of the effective Hamiltonian LOY model describing the $%
K^{0}\rightleftharpoons \overline{K}^{0}$ oscillations between a particle
and its antiparticle, we must define the velocity operator on the $\left[
\left| K^{0}\right\rangle ,\left| \overline{K}^{0}\right\rangle \right] $
basis as 
\begin{eqnarray}
\left. \frac{d\widehat{x}}{d\tau }\right| _{LOY} &\equiv &\left\langle 
\overline{s}\right| \frac{d\widehat{x}}{d\tau }\left| \overline{s}%
\right\rangle \left| K^{0}\right\rangle \left\langle K^{0}\right|
-\left\langle s\right| \frac{d\widehat{x}}{d\tau }\left| s\right\rangle
\left| \overline{K}^{0}\right\rangle \left\langle \overline{K}^{0}\right|  
\nonumber \\
&&-\left\langle s\right| \frac{d\widehat{x}}{d\tau }\left| \overline{s}%
\right\rangle \left| \overline{K}^{0}\right\rangle \left\langle K^{0}\right|
+\left\langle \overline{s}\right| \frac{d\widehat{x}}{d\tau }\left|
s\right\rangle \left| K^{0}\right\rangle \left\langle \overline{K}%
^{0}\right|   \nonumber \\
&&  \label{dx}
\end{eqnarray}
in order to describe a particle and its antiparticle according to the
Feynman prescription: antiparticles are just particles propagating backward
in time.

We complete (\ref{km2}) and write the kaons effective Hamiltonian, without
decay, but near the surface of earth 
\begin{eqnarray}
j\hbar \frac{d\left| \Psi \right\rangle }{d\tau } &=&\left( mc^{2}\widehat{I}%
-\widehat{\delta m}c^{2}\right) \cdot \left| \Psi \right\rangle   \nonumber
\\
&&-mg\left( R_{\oplus }-X\right) \widehat{I}\left| \Psi \right\rangle +mg\ 
\widehat{x}\cdot \left| \Psi \right\rangle \text{,}  \label{h1}
\end{eqnarray}
where we have dropped the $_{LOY}$ index of $\ \widehat{x}$ given by (\ref
{xs}). Then, we perform a change of variables $\Psi $ $\rightarrow $ $\Phi $
to eliminate the common phases of $a$ and $b$ associated with the inertial
mass and the small gravitational potential energy due to the average
position $R_{\oplus }+X$, 
\begin{eqnarray}
\left| \Psi \right\rangle  &=&\left| \Phi \right\rangle \exp \left( -j\frac{%
mc^{2}}{\hbar }\tau \right)   \nonumber \\
&&\times \exp j\frac{mg}{\hbar }\left( R_{\oplus }\int_{0}^{\tau
}du-\int_{0}^{\tau }X\left( u\right) du\right) \text{.}  \label{pha}
\end{eqnarray}
This new representation $\left| \Phi \right\rangle $ of the kaon state
fulfils 
\begin{equation}
j\hbar \frac{d\left| \Phi \right\rangle }{d\tau }=-\widehat{\delta m}%
c^{2}\cdot \left| \Phi \right\rangle +mg\ \widehat{x}\cdot \left| \Phi
\right\rangle \text{.}  \label{sch1}
\end{equation}
The second phase $mgR_{\oplus }\tau /\hbar $ on the right hand side of (\ref
{pha}) is not observable in experiments because it is very small ($%
mgR_{\oplus }$ $\sim 10^{-9}mc^{2}$) in front of the main phase $mc^{2}\tau
/\hbar $, and add up in the same way to both $K_{2}$ and $K_{1}$ energies so
that interferometry is impossible. The third phase $mg\int Xd\tau /\hbar $
on the right hand side of \ (\ref{pha}) is even smaller ($X\sim
10^{-7}R_{\oplus }$). Both gravity induced phase in (\ref{pha}) are
negligible in front of $mc^{2}\tau /\hbar $ and unobservable with
interferometric experiments. We can neglect the undectable gravity induced
phase of (\ref{pha}) and restrict the expression to $mc^{2}\tau /\hbar $, so
we consider that the solution $\left| \Phi \right\rangle $ of (\ref{sch1})
is related to $\left| \Psi \right\rangle $ through 
\begin{equation}
\left| \Phi \right\rangle =\left| \Psi \right\rangle \exp j\frac{mc^{2}}{%
\hbar }\tau =u\left( \tau \right) \left| K^{0}\right\rangle +v\left( \tau
\right) \left| \overline{K}^{0}\right\rangle \text{.}  \label{cov}
\end{equation}

The method used below to solve (\ref{sch1}) is based on a classical small
parameter expansion with the small parameter $\hbar g/\delta mc^{3}\sim
10^{-17}$. Then, in section V, this perturbative scheme will lead to a
classical 2 time scales problem. In order to describe the influence of 
earth's gravity on the slow $K^{0}/%
\overline{K}^{0}$ oscillation at frequency $\delta mc^{2}/\hbar $ 
we will average out the fast internal motion
(strange quarks zitterbewegung) at a frequency of the order of $mc^{2}/\hbar 
$ (or 4 times smaller if we take the mass of the strange quark). 
The ordering between these 2 time scales is $10^{-14}-10^{-15}$ thus the
classical averaging method is fully justified \cite{22,23}.

The Compton wavelength of the kaon $\lambda _{C}$ ($\hbar /mc\sim 4\times
10^{-16}$ m) provides an approximate size of the $s/\overline{s}$ quark
matrix elements $\left| \left\langle {}\right| \widehat{x}\left|
{}\right\rangle \right| $ in (\ref{xs}) because quarks are
bound states inside the volume of a kaon. The very small numerical value of
the energy $mg\lambda _{C}=\hbar g/c\sim 2.1\times 10^{-23}$ eV in front of $%
\delta mc^{2}\sim 1.7\times 10^{-6}$ eV leads to the occurrence of a very
strong ordering fulfilled by the four matrix elements $mg\left| \left\langle
{}\right| \widehat{x}\left| {}\right\rangle \right| \sim mg\lambda _{C}\ll
\delta mc^{2}$ in (\ref{sch1}). 

This very strong ordering allows to
consider a perturbative expansion scheme to solve (\ref{sch1}) with the
small expansion parameter $\hbar g/\delta mc^{3}\sim 1.2\times 10^{-17}$,
\begin{equation}
\left| \Phi \right\rangle =\left| \Phi _{0}\right\rangle +\left| \Phi
_{1}\right\rangle +O\left[ \left( \hbar g/\delta mc^{3}\right) ^{2}\right] 
\text{,}  \label{exp}
\end{equation}
where we assume the ordering $\left| \Phi _{0}\right\rangle \sim O\left[
\left( \hbar g/\delta mc^{3}\right) ^{0}\right] $ and $\left| \Phi
_{1}\right\rangle \sim O\left[ \left( \hbar g/\delta mc^{3}\right)
^{1}\right] $. With this expansion (\ref{exp}) the Schr\"{o}dinger equation (%
\ref{sch1}) becomes
\begin{eqnarray}
j\hbar \frac{d\left| \Phi _{0}\right\rangle }{d\tau } &=&-\widehat{\delta m}%
c^{2}\cdot \left| \Phi _{0}\right\rangle \text{,}  \label{eq1} \\
j\hbar \frac{d\left| \Phi _{1}\right\rangle }{d\tau } &=&-\widehat{\delta m}%
c^{2}\cdot \left| \Phi _{1}\right\rangle +mg\ \widehat{x}\cdot \left| \Phi
_{0}\right\rangle \text{.}  \label{eq2}
\end{eqnarray}
The second equation (\ref{eq2}) describes gravity induced CPV. To reveal the
dominant secular contribution of earth's gravity we introduce the inverse of
the operator $\widehat{\delta m}$: $\widehat{\delta m}^{-1}$ $=\widehat{%
\delta m}/\delta m^{2}$ and then use this operator and (\ref{eq1}) to
rewrite (\ref{eq2}) 
\begin{equation}
j\hbar \frac{d\left| \Phi _{1}\right\rangle }{d\tau }=-\widehat{\delta m}%
c^{2}\cdot \left| \Phi _{1}\right\rangle -j\frac{mg\hbar }{c^{2}}\left( \ 
\widehat{x}\cdot \frac{\widehat{\delta m}}{\delta m^{2}}\right) \cdot \frac{%
d\left| \Phi _{0}\right\rangle }{d\tau }\text{.}  \label{eq3}
\end{equation}
We define the small parameter $\kappa $ as: 
\begin{equation}
\kappa =\frac{mg\hbar }{2\delta m^{2}c^{3}}=1.7\times 10^{-3}\text{.}
\end{equation}
The equations (\ref{eq1},\ref{eq2}) are recast in a form which will appear
convenient in section V 
\begin{eqnarray}
j\frac{\hbar }{c^{2}}\frac{d\left| \Phi _{0}\right\rangle }{d\tau } &=&-%
\widehat{\delta m}\cdot \left| \Phi _{0}\right\rangle \text{,}  \nonumber \\
j\frac{\hbar }{c^{2}}\frac{d\left| \Phi _{1}\right\rangle }{d\tau } &=&-%
\widehat{\delta m}\cdot \left| \Phi _{1}\right\rangle +2j\frac{\kappa }{c}%
\left( \left. \frac{d\widehat{x}}{d\tau }\right| _{LOY}\cdot \widehat{\delta m}\right) \cdot
\left| \Phi _{0}\right\rangle   \nonumber \\
&&-2j\frac{\kappa }{c}\frac{d}{d\tau }\left[ \left( \ \left. \widehat{x}\right| _{LOY}\cdot 
\widehat{\delta m}\right) \cdot \left| \Phi _{0}\right\rangle \right] \text{.%
}  \label{eq5}
\end{eqnarray}
In doing so we have split an $O\left[ \left( \hbar g/\delta mc^{3}\right)
^{1}\right] $ term into the sum of two terms in (\ref{eq5})
displaying a different ordering. The advantage of this splitting of 
$\ \left. \widehat{x}\right| _{LOY}\cdot \widehat{\delta m}\cdot d\left| \Phi_{0}\right\rangle/{d\tau } $ 
in (\ref{eq3}) into two terms in (\ref{eq5}) is that, 
as we shall see in section V, it separates a fast oscillating term from a secular one.

To make further progress and to identify the dominant contribution of
gravity to kaons dynamics on earth we have to analyze now the behavior of
the operators $\ \left. \widehat{x}\right| _{LOY}$ (\ref{xs}) and $\left. d\ 
\widehat{x}/d\tau \right| _{LOY}$ (\ref{dx})\ involved in (\ref{eq5}).

\section{Velocity matrix elements of confined quarks: zitterbewegung}

The $s$ and $\overline{s}$ quarks are described by two $4$ components Dirac
spinors $\left| s\right\rangle $ and $\left| \overline{s}\right\rangle $
functions of the kaon's rest frame proper time $\tau $. The effective\
confining potential inside the kaon is noted $U\left( {\bf x}\right) $. This
unknown scalar potential $U$ describes the mean field strong interactions
ultimately ensuring quarks confinement in the kaon. 

The Dirac
Hamiltonians, $\widehat{H}_{S}$ and $\widehat{H}_{\overline{S}}$, and
equations are constructed on the basis of the interpretation of Feynman's
propagator theory where antiparticles are viewed as particles propagating
backward in time
\begin{eqnarray}
j\hbar d\left| s\right\rangle /d\tau  &=&\widehat{H}_{S}\left|
s\right\rangle \text{, }  \label{dirac2} \\
-j\hbar d\left| \overline{s}\right\rangle /d\tau  &=&\widehat{H}_{\overline{S%
}}\left| \overline{s}\right\rangle \text{.}  \label{dirac3}
\end{eqnarray}
We adopt this point of view because we want to describe all the particles with 
a positive mass in a way similar to the LOY model. Using a single Dirac equation, 
to describe a pair fermion-antifermion, would 
lead to the same final results, but would require the introduction of the $C$ operator 
and an interpretation of negative masses.

In the spinors Hilbert space the effective mean field Hamiltonian of the $s$
quark $\widehat{H}_{S}\left( {\bf x},{\bf p}\right) $, and of the $\overline{%
s}$ quark $\widehat{H}_{\overline{S}}\left( {\bf x},{\bf p}\right) $, are
represented by $4\times 4$ matrices 
\begin{eqnarray}
    \widehat{H}_{S} &=&c{\boldsymbol \alpha }\cdot {\bf p}+U\left( {\bf x}\right) \beta
    +m_{s}c^{2}\beta \text{,}  \label{HD2} \\
    \widehat{H}_{\overline{S}} &=&-c{\boldsymbol \alpha }\cdot {\bf p}+U\left( {\bf x}%
    \right) \beta +m_{\overline{s}}c^{2}\beta \text{.}  \label{HD3}
    \end{eqnarray} 
    where we have introduced the set of $4\times 4$ matrices: ${\boldsymbol \alpha }$ $=$
    $\left( \alpha _{x},\alpha _{y},\alpha _{z}\right) $ and $\beta $ \cite{18},
    expressed in terms of the Pauli matrices ${\boldsymbol\sigma }$ $=$ $\left( \sigma
    _{x},\sigma _{y},\sigma _{z}\right) $ and the $2\times 2$ identity $I$ as 
    \begin{equation}
    {\boldsymbol \alpha }=\left( 
    \begin{array}{cc}
    0 & {\boldsymbol\sigma } \\ 
    {\boldsymbol\sigma } & 0
    \end{array}
    \right) \text{, }\beta =\left( 
    \begin{array}{cc}
    I & 0 \\ 
    0 & -I
    \end{array}
    \right) \text{. }  \label{rpz}
    \end{equation}
The position and momentum operators ${\bf x}$ and ${\bf p}$ fulfill the
canonical commutations rules $\left[ x_{i},p_{j}\right] $ = $j\hbar \delta
_{j}^{i}$. One of the puzzling property of Dirac's equations (\ref{dirac2})
and (\ref{dirac3}) is the following expression of the Heisenberg velocity
operator
\begin{equation}
    j\hbar \frac{d{\bf x}}{d\tau }=\left[ {\bf x},\widehat{H}\right] =j\hbar c%
    {\boldsymbol \alpha }  \label{zzz}.
    \end{equation}
Here $\widehat{H}$ is either (\ref{HD2}) or (\ref{HD3}). This result (\ref
{zzz}) is independent of the charge, mass and confining potential $U$ of the
quarks inside the kaons and is part of the various nonintuitive behaviors of
relativistic spin $1/2$ particles named {\it zitterbewegung }\cite{18}. 
The
term {\it zitterbewegung\ }as used here is given wide meaning. It means all
the unusual properties associated with the velocity of spin $1/2$
particles/antiparticles, either free or bound.

Without loss of generality, we take the $x$ direction of the Dirac
representation (\ref{rpz}) as the vertical direction and $d\widehat{x}/d\tau
=c\alpha _{x}$. The dynamics along this $x$ vertical direction in the kaon
rest frame is independent of the $y$ and $z$ dynamics which are not coupled
to gravity. 

We consider an effective confining potential $U\left( x\right) $
and take $p_{y}=p_{z}=0$. We look for two strange quark energy eigenstates
with $E>0$: for $\left| s\right\rangle $ a solution of (\ref{dirac2}) of the
type $\left\langle x\right. \left| s\right\rangle \exp -jEt/\hbar $ and for $%
\left| \overline{s}\right\rangle $ a solution of (\ref{dirac3}) of the type $%
\left\langle x\right. \left| \overline{s}\right\rangle \exp -jEt/\hbar $.

The $8$ first order differential equations (\ref{dirac2}) and (\ref{dirac3})
are in fact $4$ times the repetition of a similar reduced set of $2$
equations describing the coupling between two unknown wave-functions $\chi
_{E}\left( x\right) $ and $\zeta _{E}\left( x\right) $ such that 
\begin{eqnarray}
j\hbar c\partial \zeta _{E}/\partial x+\left[ E+U
+m_{s}c^{2}\right] \chi _{E}  &=&0\text{,}  \label{d1} \\
j\hbar c\partial \chi_{E}/\partial x+\left[ E-U
-m_{s}c^{2}\right] \zeta _{E} &=&0\text{.}  \label{d2}
\end{eqnarray}
We consider the two localized solutions of (\ref{d1}) and (\ref{d2}) such
that $\zeta _{E}\left( \pm \infty \right) $ = $\chi _{E}$ $\left( \pm \infty
\right) $ = $0$ and normalize them according to 
\begin{equation}
\int \zeta _{E}^{*}\zeta _{E}dx=\int \chi _{E}^{*}\chi _{E}dx=1\text{.}
\label{pop2}
\end{equation}
From (\ref{d1}) and (\ref{d2}) we obtain $\int \zeta _{E}\chi
_{E}^{*}dx=\int \chi _{E}\zeta _{E}^{*}dx$ and $\partial \left( \chi
_{E}\zeta _{E}^{*}+\zeta _{E}\chi _{E}^{*}\right) /\partial x=0$ which imply
the additional orthogonality property: 
\begin{equation}
\int \zeta _{E}^{*}\chi _{E}dx=\int \zeta _{E}\chi _{E}^{*}dx=0\text{.}
\label{pop}
\end{equation}
In the following we drop the energy eigenvalue index $E$. 

These two
functions $\zeta $ and $\chi $ provide the general solutions
of the $8$ differential equations (\ref{dirac2}, \ref{dirac3}) 
describing  strange quarks confined in the same mean field
\begin{eqnarray}
\left\langle s\right. \left| x\right\rangle  &=&\left[ \lambda \chi \left(
x\right) ,\mu \chi \left( x\right) ,\mu \zeta \left( x\right) ,\lambda \zeta
\left( x\right) \right] /N\text{, } \\
\left\langle \overline{s}\right. \left| x\right\rangle  &=&\left[ \lambda
\zeta \left( x\right) ,\mu \zeta \left( x\right) ,\mu \chi \left( x\right)
,\lambda \chi \left( x\right) \right] /N\text{.}
\end{eqnarray}
The two complex numbers $\left( \lambda ,\mu \right) $ specify the spin
state and $N=\sqrt{2\left( \lambda \lambda ^{*}+\mu
\mu ^{*}\right) }$ ensures the normalization: $\left\langle s\right. \left|
s\right\rangle =1$ and $\left\langle \overline{s}\right. \left| \overline{s}%
\right\rangle =1$. The matrix elements $\left\langle q\right| \alpha
_{x}\left| q^{\prime }\right\rangle $ = $\int dx\left\langle q\right. \left|
x\right\rangle .\alpha _{x}.\left\langle x\right. \left| q^{\prime
}\right\rangle $ and the use of (\ref{pop}, \ref{pop2}) gives 
\begin{eqnarray}
\left\langle \overline{s}\right| \alpha _{x}\left| \overline{s}\right\rangle
&=&\left\langle s\right| \alpha _{x}\left| s\right\rangle =0\text{,} 
\nonumber \\
\left\langle s\right| \alpha _{x}\left| \overline{s}\right\rangle 
&=&\left\langle \overline{s}\right| \alpha _{x}\left| s\right\rangle =1\text{%
.}  \label{zzz1}
\end{eqnarray}
This set of relations is independent of the quark mass, charge, confining
potential and energy eigenvalue $E$ (zitterbewegung effect). The definition
of the velocity operator (\ref{dx}), associated with the
particle-antiparticle LOY model,  becomes
\begin{equation}
\left. \frac{d\widehat{x}}{d\tau }\right| _{LOY}=\left[ 
\begin{array}{cc}
0 & c \\ 
-c & 0
\end{array}
\right] \text{,}  \label{ccc1}
\end{equation}
in the kaon Hilbert space $\left[ \left| K^{0}\right\rangle ,\left| 
\overline{K}^{0}\right\rangle \right] $. 

In the study \cite{19} based on two
coupled Klein-Gordon equations on a Schwarzschild metric we obtained this
result (\ref{zzz1}) with a different model so this zitterbewegung behavior (%
\ref{zzz1}) is model independent.

\section{Interpretation of CP violation neutral kaons experiments}

Collecting the results (\ref{ccc1}) and (\ref{eq5}) and introducing the
amplitudes ($u_{0},v_{0}$) and ($u_{1},v_{1}$) of the kaon state (\ref{exp}) 
\begin{equation}
\left| \Phi _{0/1}\right\rangle =u_{0/1}\left( \tau \right) \left|
K^{0}\right\rangle +v_{0/1}\left( \tau \right) \left| \overline{K}%
^{0}\right\rangle \text{,}  \label{cov2}
\end{equation}
the impact of gravity on neutral kaons oscillations is described by (\ref{eq5}) which can be written 
\begin{eqnarray}
j\frac{\hbar }{\delta mc^{2}}\frac{d}{d\tau }\left[ 
\begin{array}{l}
u_{0} \\ 
v_{0}
\end{array}
\right]  &=&-\left[ 
\begin{array}{cc}
0 & 1 \\ 
1 & 0
\end{array}
\right] \cdot \left[ 
\begin{array}{l}
u_{0} \\ 
v_{0}
\end{array}
\right] \text{,}  \nonumber \\
j\frac{\hbar }{\delta mc^{2}}\frac{d}{d\tau }\left[ 
\begin{array}{l}
u_{1} \\ 
v_{1}
\end{array}
\right]  &=&-\left[ 
\begin{array}{cc}
0 & 1 \\ 
1 & 0
\end{array}
\right] \cdot \left[ 
\begin{array}{l}
u_{1} \\ 
v_{1}
\end{array}
\right]   \nonumber \\
&&+2j\kappa \left[ 
\begin{array}{cc}
1 & 0 \\ 
0 & -1
\end{array}
\right] \cdot \left[ 
\begin{array}{l}
u_{0} \\ 
v_{0}
\end{array}
\right]   \label{eq66} \\
&&-2j\kappa \frac{d}{d\tau }\left( \frac{\ \widehat{x}\left( \tau \right) }{c%
}\cdot \frac{\widehat{\delta m}}{\delta m}\cdot \left[ 
\begin{array}{l}
u_{0} \\ 
v_{0}
\end{array}
\right] \right) \text{.}  \nonumber
\end{eqnarray}
We can not express $\left. \widehat{x}\left( \tau \right) \right| _{LOY}$ $\ 
$because we do not know $U\left( x\right) $, but we can evaluate the
behavior of the last term of (\ref{eq66}). As the extent of the confining
effective potential $U$ is of the order of $\lambda _{C}=\hbar /mc$, the
typical order of magnitude of the matrix elements of the type  $\left\langle 
\overline{q}\right| \widehat{x}\left( \tau \right) \left| q\right\rangle $
or $\left\langle q\right| \widehat{x}\left( \tau \right) \left|
q\right\rangle $  will be in between zero and  $\ \lambda _{C}=\hbar /mc$.
Considering the time independent part of  $\left\langle q\right| \widehat{x}%
\left( \tau \right) \left| q\right\rangle $ or $\left\langle \overline{q}%
\right| \widehat{x}\left( \tau \right) \left| q\right\rangle $ (if any), and
taking the size upper bound $\hbar /mc$, the last term of (\ref{eq66})
 will scale as $\kappa \delta m/m\sim 10^{-15}\kappa $. Thus the constant part
of the matrix elements can be safely neglected. The matrix elements of the
type $\left\langle \overline{q}\right| \widehat{x}\left( \tau \right) \left|
q\right\rangle $ or $\left\langle q\right| \widehat{x}\left( \tau \right)
\left| q\right\rangle $ might also contain a high frequency part  at
the zitterbewegung frequency $c/\lambda _{C}=mc^{2}/\hbar $, which is to be
compared to the $K^{0}\rightleftharpoons \overline{K}^{0}$ mixing frequency $%
\delta mc^{2}/\hbar \sim 10^{-15}mc^{2}/\hbar $. 

Two different time scales
are thus involved in (\ref{eq66}). The first one is a fast zitterbewegung
oscillation of the confined strange quark, with a typical frequency $%
c/\lambda _{C}=mc^{2}/\hbar $. The slow time scale is associated with the $%
K^{0}\rightleftharpoons \overline{K}^{0}$ mixing oscillation at frequency $%
\delta mc^{2}/\hbar $. The ordering between these slow and fast time scales
is given by the ratio $10^{-14}-10^{-15}$.

The value of the ratio between the two frequencies allows to
average out safely the fast oscillation in the $\left( u_{1},v_{1}\right) $
equation (\ref{eq66}) to describe the slow dynamics of the $\Delta S=2$
mixing modified by gravity. 

We consider one period $\theta $ of the
(unknown) periodic functions $\left\langle {}\right| \widehat{x}\left(
t\right) \left| {}\right\rangle $, $\theta $ is such that $\hbar /mc^{2}\sim
\theta \ll $ $\hbar /\delta mc^{2}$. We apply the averaging operator $%
\widehat{A}_{\theta }\equiv $\ $\int_{\tau }^{\tau +\theta }dt/\theta $ on
both side of (\ref{eq66}) to average the high
frequency components. For any slow function $h\left( t\right) $: $\widehat{A}%
_{\theta }.h=h\left( \tau \right) $ and $\widehat{A}_{\theta }.dh/dt=$ $%
dh/d\tau $. The high frequency part of $\left( u_{1},v_{1}\right) $ and $%
d\left( u_{1},v_{1}\right) /d\tau $, driven by the high frequancy part of
 $\left\langle {}\right| \widehat{x}\left( t\right) \left| {}\right\rangle $, 
 average to zero and the last term 
\begin{equation}
\int_{\tau }^{\tau +\theta }\frac{d}{dt}\ \widehat{x}\cdot \widehat{\delta m}%
\cdot \left( 
\begin{array}{l}
u_{0} \\ 
v_{0}
\end{array}
\right) dt=\ \left. \widehat{x}\cdot \widehat{\delta m}\cdot \left( 
\begin{array}{l}
u_{0} \\ 
v_{0}
\end{array}
\right) \right| _{\tau }^{\tau +\theta }=0\text{,}
\end{equation}
cancels as $\left( u_{0},v_{0}\right) $ does not change during the time $%
\theta $. We have just used here the usual averaging method of classical
oscillators theory to separate a slow dynamics from a fast  one \cite{22,23}.

The $\left| \Phi \right\rangle $ amplitudes  $\left( u,v\right) $ = $\left(
u_{0},v_{0}\right) $ + $\left( u_{1},v_{1}\right) $ fulfill
\begin{equation}
j\frac{\hbar }{c^{2}}\frac{d}{d\tau }\left[ 
\begin{array}{l}
u \\ 
v
\end{array}
\right] =\delta m\left[ 
\begin{array}{cc}
2j\kappa  & -1 \\ 
-1 & -2j\kappa 
\end{array}
\right] \cdot \left[ 
\begin{array}{l}
u \\ 
v
\end{array}
\right] \text{.}
\end{equation}
to the lowest order in $\kappa $. Then, back to the kaons $\Psi $ amplitudes 
$\left( a,b\right) $ (\ref{ampK}) from the  kaons $\Phi $ amplitudes $\left(
u,v\right) $ (\ref{cov}), the effective Hamiltonian modified by gravity
is 
\begin{equation}
j\frac{\hbar }{c^{2}}\frac{d}{d\tau }\left[ 
\begin{array}{l}
a \\ 
b
\end{array}
\right] =\left[ 
\begin{array}{cc}
m+2j\kappa \delta m & -\delta m \\ 
-\delta m & m-2j\kappa \delta m
\end{array}
\right] \cdot \left[ 
\begin{array}{l}
a \\ 
b
\end{array}
\right] \text{.}  \label{finalx}
\end{equation}
The two stables states, $\left| \widetilde{K}_{L}\right\rangle $ and $\left| 
\widetilde{K}_{S}\right\rangle $, solutions of (\ref{finalx}) evolve with
time according to 
\begin{equation}
\left| \widetilde{K}_{L/S}\right\rangle =\exp -j\left( m\pm \delta m\right) 
\frac{c^{2}}{\hbar }\tau \left| K_{L/S}^{\oplus }\right\rangle \text{,}
\label{ls}
\end{equation}
where the two eigenvectors, $K_{L}^{\oplus }$ and $K_{S}^{\oplus }$, are
\begin{eqnarray}
\left| K_{L/S}^{\oplus }\right\rangle  &=&\left[ \left( 1\pm j\kappa \right)
\left| K^{0}\right\rangle \mp \left( 1\mp j\kappa \right) \left| \overline{K}%
^{0}\right\rangle \right] /\sqrt{2}\text{,}  \nonumber \\
\left| K_{L/S}^{\oplus }\right\rangle  &=&\left| K_{2/1}\right\rangle \pm
j\kappa \left| K_{1/2}\right\rangle \text{.}  \label{kl2}
\end{eqnarray}
Small $O\left[ \kappa ^{2}\right] $ $\sim 10^{-6}$ terms are neglected and
these eigenvectors are normalized to one with this accuracy. 

The very first
consequence of (\ref{kl2}) is the prediction of the amplitude of the
indirect violation 
\begin{equation}
\frac{\left\langle K_{S}^{\oplus }\right. \left| K_{L}^{\oplus
}\right\rangle }{2}=j\kappa \text{.}  \label{keke}
\end{equation}

The two superpositions (\ref{kl2}) are the (stable) kaon energy eigenstates
on earth, but they are no longer $CP$ eigenstates. Gravity induced CPV is
thus demonstrated. But this is a $T$ conservative CPV
with a diagonal perturbation $m\rightarrow m\pm 2j\kappa \delta m$ in (\ref
{finalx}) with respect to (\ref{km2}) and (\ref{m123}), while the usual $CPT$
conservative CPV non diagonal perturbation $\delta m\rightarrow \delta m\pm
2\varepsilon \delta m$ gives the following kaons energy eigenstates
\begin{eqnarray}
\left| K_{L/S}\right\rangle  &=&\left[ \left( 1+\varepsilon \right) \left|
K^{0}\right\rangle \mp \left( 1-\varepsilon \right) \left| \overline{K}%
^{0}\right\rangle \right] /\sqrt{2}\text{,}  \nonumber \\
\left| K_{L/S}\right\rangle  &=&\left| K_{2/1}\right\rangle +\varepsilon
\left| K_{1/2}\right\rangle \text{,}  \label{epg2}
\end{eqnarray}
considered usually to interpret CPV experiments rather than (\ref{kl2}).

The previous results (\ref{kl2}) and (\ref{ls}) do not predict the impact of
gravity on CPV (unstable) kaons experiments, but the impact of gravity on
stable kaons oscillations. To be able to predict the measured impact of
gravity on CPV experiments we take into account now the $K_{S}$ finite
lifetime and neglect the finite lifetime of $K_{L}$ which is 600 times
larger than the $K_{S}$ one.

There is a difference between ({\it i}) the
usual $K_{S}$ regeneration from $K_{L}$ through the interaction with a
material environment and ({\it ii}) gravitational regeneration associated
with (\ref{keke}). Gravitational regeneration is continuous and create a
particular quantum state as a result of the balance between the decay and
the regeneration of $K_{S}$, as opposed to usual regeneration which is just
an initial value problem starting at the exit of the regenerator.

The kaons observed in CPV experiments are not the stable eigenstates (\ref
{ls}). At the very beginning (kaon production) of a CPV experiment there is
an equal amount of $K_{1}$ and $K_{2}$. Then the short-lived kaon decay away
rapidly. 

After the decay of this initial short-lived kaon content, the
kaons are in a particular quantum state associated with a
continuous balance between $K_{S}$ regeneration and decay. The long-lived state
should match the $\left| K_{L}^{\oplus }\right\rangle $ $=\left|
K_{2}\right\rangle +j\kappa \left| K_{1}\right\rangle $ (\ref{kl2}) state,
but its $j\kappa \left| K_{1}\right\rangle $ component decay away very
rapidly. To remain an energy eigenstate the $K_{L}$ regenerate continuously
this small $j\kappa \left| K_{1}\right\rangle $ component. The result is
that $\Psi _{\exp }$, the long-lived kaon state observed in CPV\
experiments, is to be described by the solution of a Schr\"{o}dinger
equation including both regeneration and decay.

We introduce $A$ and $B$ the amplitudes of the experimental kaon state $\Psi
_{\exp }$ observed in CPV\ experiments 
\begin{equation}
\left| \Psi _{\exp }\right\rangle =A\left( \tau \right) \left|
K_{1}\right\rangle +B\left( \tau \right) \left| K_{2}\right\rangle \text{.}
\end{equation}
We assume that the $K_{2}$ component is stable and that the depletion of its amplitude
associated with gravitational regeneration is negligible. 

The gravity
induced regeneration rate is given by the transition amplitude per unit time 
$\omega $ =  $d\left\langle \widetilde{K}_{L}\right. \left| \widetilde{K}%
_{S}\right\rangle /d\tau $ whith the kaons states $\left| \widetilde{K}%
_{S/L}\right\rangle $ defined  by (\ref{ls})
\begin{equation}
\omega =2j\frac{\delta mc^{2}}{\hbar }\left\langle K_{L}^{\oplus }\right.
\left| K_{S}^{\oplus }\right\rangle \exp \left( 2j\delta m\frac{c^{2}}{\hbar 
}\tau \right) \text{.}  \label{rate2}
\end{equation}
The short-lived kaon decays will be described as an irreversible process
within the framework of the Weisskopf-Wigner approximation \cite{20}, with a
rate  $\left( \Gamma -\delta \Gamma \right) /2$ $\approx -\delta \Gamma
\approx \Gamma $ (\ref{m123}). We can anticipate that $T$ \ will appear to
be violated as a result of this description of the decay as an irreversible process.

The amplitude $A$ of the short-lived kaon fulfils the Schr\"{o}dinger
equation 
\begin{equation}
j\hbar \frac{dA}{d\tau }=\left( m-\delta m\right) c^{2}A-j\hbar \Gamma
A+j\hbar \omega B\text{.}  \label{sh56}
\end{equation}

The amplitude $B$ of a  driving long-lived kaon normalized state is such that $BB^{*}=1$
(no depletion), so we take the amplitude 
\begin{equation}
B=\exp -j\left( m+\delta m\right) \frac{c^{2}}{\hbar }\tau \text{.}
\end{equation}

With this driving amplitude the solution $A$ of  Schr\"{o}dinger's equation (%
\ref{sh56}) is 
\begin{equation}
A=2\left( \frac{\delta mc^{2}}{\hbar \Gamma /2}\right) \kappa \exp -j\left(
m-\delta m\right) \frac{c^{2}}{\hbar }\tau \text{,}
\end{equation}
where we have used (\ref{rate2}) with $\left\langle K_{L}^{\oplus }\right.
\left| K_{S}^{\oplus }\right\rangle =  -2j\kappa $ (\ref{keke}).\ 

The observation of a long-lived kaon on earth in CPV experiments is in fact the
observation of the state $\left| \Psi _{\exp }\right\rangle $ 
\begin{eqnarray}
\left| \Psi _{\exp }\right\rangle  &=&\exp -j\left( m+\delta m\right) \frac{%
c^{2}}{\hbar }\tau \left| K_{2}\right\rangle   \label{exp1} \\
&&+2\left( \frac{\delta mc^{2}}{\hbar \Gamma /2}\right) \kappa \exp -j\left(
m-\delta m\right) \frac{c^{2}}{\hbar }\tau \left| K_{1}\right\rangle \text{,}
\nonumber
\end{eqnarray}
where $O\left[ \kappa ^{2}\right] $ terms are neglected. The measurment of
the overlap $\left\langle K_{S}^{\oplus }\right. \left| K_{L}^{\oplus
}\right\rangle $ is a measurment where $\left| K_{L}^{\oplus }\right\rangle $
is in fact $\left| \Psi _{\exp }\right\rangle $ (\ref{exp1}) and $%
\left\langle K_{S}^{\oplus }\right| $ is  in fact\ $\left\langle
K_{1}\right| \exp j\left( m-\delta m\right) c^{2}\tau /\hbar $. The impact
of the short-lived kaon continuous decay and regeneration is that the indirect
violation  parameter (\ref{keke}) observed on earth   becomes
\begin{equation}
\frac{\left\langle K_{S}^{\oplus }\right. \left| K_{L}^{\oplus
}\right\rangle _{\exp }}{2}=\left( \frac{\delta mc^{2}}{\hbar \Gamma /2}%
\right) \kappa =1.6\times 10^{-3}\text{.}  \label{ep}
\end{equation}

As expected $T$ conservation no longer holds because $\left\langle
K_{L}^{\oplus }\right. \left| K_{S}^{\oplus }\right\rangle _{\exp }$ is real 
\cite{4}. We conclude that the measured CPV parameter is a real number
associated with $CPT$ conservation and $T$ \ violation (due to the
irreversibility of decays) despite the fact that the fundamental{\it \ }$%
\left\langle K_{L}^{\oplus }\right. \left| K_{S}^{\oplus }\right\rangle $ is
an imaginary number associated with $CPT$ violation (due to gravity) and $T$
conservation. 

The experimental result  (\ref{ep}) has led in the past to interpret the experiments 
with  two CPV states of the type (\ref{epg2}): $\left| K_{L/S}\right\rangle =\left| K_{2/1}\right\rangle 
+ \varepsilon
\left| K_{1/2}\right\rangle $,  with the value
\begin{equation}
\mathop{\rm Re}%
\varepsilon =1.6\times 10^{-3}\,  \label{eepp}
\end{equation}
as these states (\ref{epg2}) give $\left\langle K_{L}\right. \left|
K_{S}\right\rangle =2%
\mathop{\rm Re}%
\varepsilon $. But this interpretation hides two subtle points: ({\it i}) $CPT$ is violated by
gravity with $T$ conservation, and ({\it ii}) the observed $T$ violation stems from
the irreversible decays of $K_{S}$. 

As the $d/\overline{d}$ quark dynamics
has been neglected we can predict the gravity induced $CP$ violation with an
accuracy of the order of 4\% and this theoretical expression of $%
\mathop{\rm Re}%
\left( \varepsilon \right) $ (\ref{eepp}) is in agreement with the
experimental data with this precision \cite{21}.\ 

We consider now both $K_{L}$ finite lifetime and $K_{S}$ finite lifetime to
calculate the observed $\arg \varepsilon $. We use here Bell-Steinberger's
unitarity relations \cite{24}. The depletion of a kaon state $\left| \Psi
\right\rangle $ is described by the set of amplitudes $\left\langle f\right| 
{\cal T}\left| \Psi \right\rangle $ according to 
\begin{equation}
\left. \partial \left\langle \Psi \right. \left| \Psi \right\rangle
/\partial \tau \right| _{\tau =0}=-\Sigma _{f}\left| \left\langle f\right| 
{\cal T}\left| \Psi \left( \tau =0\right) \right\rangle \right| ^{2}\text{,}
\label{BS1}
\end{equation}
where $f$ \ runs over the set of all the allowed final states. 

Introducing
the inverse of the mean lifetimes, $\Gamma _{S}=\Gamma -\delta \Gamma
\approx 2\Gamma $ and $\Gamma _{L}=\Gamma +\delta \Gamma $, of $K_{1}$ and $
K_{2}$, the model (\ref{km2}) leads to 3 relations between the effective
Hamiltonian parameters ($\Gamma _{S}$, $\Gamma _{L}$, $\delta m$) and the
amplitudes $\left\langle f\right| {\cal T}\left| K_{S/L}^{\oplus
}\right\rangle $%
\begin{eqnarray}
\Gamma _{S/L} &=&\Sigma _{f}\left| \left\langle f\right| {\cal T}\left|
K_{S/L}^{\oplus }\right\rangle \right| ^{2}\text{,}  \label{bs23} \\
2j\frac{\delta mc^{2}}{\hbar }+\frac{\Gamma _{S}}{2} &=&\Sigma _{f}\frac{%
\left\langle f\right| {\cal T}\left| K_{L}^{\oplus }\right\rangle
\left\langle f\right| {\cal T}\left| K_{S}^{\oplus }\right\rangle ^{*}}{
\left\langle K_{S}^{\oplus }\right. \left| K_{L}^{\oplus }\right\rangle }
\text{.}  \label{bs5}
\end{eqnarray}
For $K_{S}$ the sum over $f$ \ is dominated by $K_{S}\rightarrow $ $2\pi $
decays (99.9\%), and more precisely by the $K_{S}$ decay to the isospin-zero
combination of $\left| \pi ^{+}\pi ^{-}\right\rangle $ and $\left| \pi
^{0}\pi ^{0}\right\rangle $ noted $\left| I_{0}\right\rangle $ (the decays
to $\left| I_{2}\right\rangle $ account for about 5\% ). Neglecting the
others decays,\ and restricting the right hand side of (\ref{bs5}) and (\ref{bs23})
to $\left| f\right\rangle =\left| I_{0}\right\rangle $, we obtain 
\begin{equation}
2j\frac{\delta mc^{2}}{\hbar }+\frac{\Gamma _{S}}{2}=\frac{\left\langle I_{0}\right| 
{\cal T}\left| K_{L}^{\oplus }\right\rangle \left\langle I_{0}\right| {\cal T
}\left| K_{S}^{\oplus }\right\rangle ^{*}}{\left\langle K_{S}^{\oplus
}\right. \left| K_{L}^{\oplus }\right\rangle }\text{.}  \label{bs7}
\end{equation}
The right hand side of (\ref{bs7}) is then expressed with the help of the
definition of the indirect violation parameter $\varepsilon $ and Eq. (\ref
{bs23}) 
\begin{equation}
\varepsilon \equiv \frac{\left\langle I_{0}\right| {\cal T}\left|
K_{L}^{\oplus }\right\rangle }{\left\langle I_{0}\right| {\cal T}\left|
K_{S}^{\oplus }\right\rangle }=\frac{\left\langle I_{0}\right| {\cal T}
\left| K_{L}^{\oplus }\right\rangle \left\langle I_{0}\right| {\cal T}\left|
K_{S}^{\oplus }\right\rangle ^{*}}{\Gamma _{S}}\text{.}
\end{equation}
This expression is then substituted in (\ref{bs7}) with the help of (\ref{ep}) to obtain the final expression 
\begin{equation}
\left( 2j\frac{\delta mc^{2}}{\hbar }+\frac{\Gamma _{S}}{2}\right) 2\kappa
\left( \frac{\delta mc^{2}}{\hbar \Gamma /2}\right) =\varepsilon \Gamma _{S}
\text{,}  \label{bs56}
\end{equation}
Thus, as $\Gamma _{S}=2\Gamma$, the argument of the measured
CPV parameter $\varepsilon $ is 
\begin{equation}
\arg \left( \varepsilon \right) =\arctan \left( \frac{2\delta mc^{2}}{\hbar
\Gamma }\right) =43.4^{\circ }\text{.}
\end{equation}
This result complete the relation (\ref{ep})  providing $
\mathop{\rm Re}
\left( \varepsilon \right) $ 
\begin{equation}
\mathop{\rm Re}%
\left( \varepsilon \right) =\left( \frac{\delta mc^{2}}{\hbar \Gamma /2}
\right) \left( \frac{mg\hbar }{2\delta m^{2}c^{3}}\right) =1.6\times 10^{-3}
\text{.}
\end{equation}

This analysis demonstrates that earth's
gravity, described by its Newtonian potential, without any speculative hypothesis,
induces a small CPV and that
the predicted indirect violation parameter $\varepsilon $ is in agreement
with the experimental measurements \cite{21}.

To interpret the measurements of the direct violation parameter $\varepsilon
^{\prime }$ we consider the $2\pi $ decays of $K_{L}$ and $K_{S}$ and the
definition of the amplitude ratio $\eta _{00}$ as a function of $\varepsilon 
$ and $\varepsilon ^{\prime }$%
\begin{equation}
\eta _{00}\equiv \frac{\left\langle \pi ^{0}\pi ^{0}\right| {\cal T}\left|
K_{L}\right\rangle }{\left\langle \pi ^{0}\pi ^{0}\right| {\cal T}\left|
K_{S}\right\rangle }\widetilde{RF}=\varepsilon -2\varepsilon ^{\prime }\text{
,}  \label{am2}
\end{equation}
where $\widetilde{RF}$ is a rephasing factor needed to define a phase
convention independent amplitude ratios $\eta _{00}$ and obtain a phase
convention invariant quantity, i.e. a meaningful physical quantity. This
factor is one under the assumption of $CPT$ conservation. The most simple 
 rephasing factors $\widetilde{RF}$ for $\eta _{00}$ is 
\begin{equation}
\widetilde{RF}=\frac{\left\langle K^{0}\right. \left| K_{S}^{\oplus
}\right\rangle }{\left\langle K^{0}\right. \left| K_{L}^{\oplus
}\right\rangle }=1-2j\kappa =1-\left\langle K_{S}^{\oplus }\right. \left|
K_{L}^{\oplus }\right\rangle \text{,}  \label{rp}
\end{equation}
where we have used the states (\ref{kl2}). But this result (\ref{rp}) does
not take into account the finite lifetime of $K_{L}$. This dissipative
effect operates a rotation in the complex plane from $j\kappa $ in (\ref
{keke}) to $\left( 2\delta mc^{2}/\hbar \Gamma \right) \kappa $ in (\ref{ep}), so that the rephasing factor with $K_{L}$ finite lifetime becomes $%
1-2\left( 2\delta mc^{2}/\hbar \Gamma \right) \kappa $. To calculate the two
amplitudes  in (\ref{am2}) some additional properties are needed. We assume
that the amplitude of $ K_{0}\rightarrow \pi ^{0}\pi
^{0}$ is equal to the amplitude of $ \overline{K}^{0}\rightarrow
\pi ^{0}\pi ^{0} $ because the interaction between a $\left( \pi
^{0}\pi ^{0}\right) $ state and a neutral kaon state, $K_{0}$ or $\overline{K
}^{0}$, can not differentiate the $K^{0}$ from the $\overline{K}^{0}$ and a
final state phase can be absorbed by a proper phase convention between the $
K_{0}$ and $\overline{K}^{0}$. Using the states (\ref{epg2}) to calculate
the ratio\ of the  amplitudes in (\ref{am2}) leads to the final expression 
\begin{equation}
\eta _{00}=\varepsilon \left[ 1-2\left( 2\delta mc^{2}/\hbar \Gamma \right)
\kappa \right] \text{.}  \label{etha}
\end{equation}
As a consequence of the definition of $\varepsilon ^{\prime }$ in (\ref{am2}) 
\begin{equation}
\mathop{\rm Re}%
\left( \frac{\varepsilon ^{\prime }}{\varepsilon }\right) =\left( \frac{
\delta mc^{2}}{\hbar \Gamma /2}\right) \kappa =1.6\times 10^{-3}\text{,}
\end{equation}
in agreement with the experimental data \cite{21}. The other consequence of
(\ref{etha}) is $\arg \left( \varepsilon \right) =\arg \left( \varepsilon
^{\prime }\right) $.

The usual {\it a priori }statements on $CPT$
conservation is to be revisited to construct the complete phase convention
independent interpretation of direct $CP$ violation measurements. The
study of the $2\pi $ decays presented here to extract $
\mathop{\rm Re}
\left( \varepsilon ^{\prime }/\varepsilon \right) $ and $\arg \varepsilon
^{\prime }/\arg \varepsilon $ $\ $is only an illustrative example of the
program of re-interpretation of the commonly assumed $CPT$ conserving
neutral kaons anomalous decays data, and more generally neutral meson
oscillations experiments within the framework of the small $CPT$ violation
induced by earth's gravity.

\section{Conclusions and perspectives}

The previous calculations can be extended to any environment where
oscillating neutral kaons experience an acceleration ${\bf a}$ as a result of
the interaction with an external field or a curved geometry. 

The amplitude
of indirect $CP$ violation as predicted by (\ref{ep}) will be the product of
three factors
\begin{equation}
\left\langle K_{S}\right. \left| K_{L}\right\rangle _{{\bf a}}=4\left( \frac{
m}{\delta m}\right) ^{2}\left( \frac{\delta mc^{2}}{\hbar \Gamma }\right)
\ \frac{\left| {\bf a}\right| \lambda _{C}}{c^{2}}\ \text{.}
\label{fac}
\end{equation}
The first two factors are associated with electroweak and strong
interactions described in (\ref{mass}), the third one is associated with the external field. 

We hope that the new gravity (or ${\bf a}$) induced CPV mechanism described by (\ref{ep}) 
(or (\ref{fac})) 
will provide the right framework to set up
cosmological evolution models predicting the strong asymmetry between the
abundance of matter and the abundance of anti-matter in our present universe.

Besides the problem of early baryogenesis, neutrinos oscillations near a
spherical massive object might be revisited to explore the impact of the
interplay between gravity and mixing.

The SM model in its present form provides the canonical framework to
understand the hierarchy, dynamics and symmetries of elementary particles in
the vicinity of earth's surface. 

The identification of gravity as the source
of CPV in kaons experiments does not contradict this model, but elucidates 
the origin of one of the CKM  matrix parameters, and re-establishes time reversal symmetry   
at the microcopic level. The consequences of a CKM matrix without CPV
should be considered as a model of elementary particles far from any massive object.

To summarize our findings: ({\it i}) the origin of $CP$ violation in neutral
kaons systems is identified as earth's gravity, ({\it ii}) the measured
values of both the direct and the indirect $CP$ violation parameters are
predicted and ({\it iii}) the right status of the symmetry breaking induced
by the interplay between gravity, quarks zitterbewegung and $\Delta
S=2$ mixing oscillations is restored: $CPT$ violation with $T$ conservation
rather than $T$ violation with $CPT$ conservation{\it . }Because experiments
are performed with $K_{S}$ generated through a balance between gravity
induced regeneration and  irreversible decay, time reversal symmetry is
broken under these experimental conditions and gravity induced CPV is
usually viewed as $CPT$ conservation with $T$ and $CP$ violations.
\begin{acknowledgments}
This study was undertaken during a sabbatical year at Princeton and completed
at University of Paris-Saclay. The financial support of a G. R. Andlinger
fellowship from Princeton University ACEE and the hospitality of Professor Nat Fisch's team 
are gratefully acknowledged.  
\end{acknowledgments}

    \end{document}